\documentstyle[preprint,aps]{revtex}

\begin{document}
\draft
\title{Two-Mode Squeezed States and Their Superposition in the Motion of Two
Trapped Ions}
\author{Hao-Sheng Zeng$^{1,2}$\thanks{%
E-mail adress: hszeng@mail.hunnu.edu.cn}, Le-Man Kuang$^{1}$ and Ke-Lin Gao$%
^{2}$}
\address{$^{1}$Department of Physics, Hunan Normal University, Hunan\\
410081, People's Republic of China\\
$^{2}$Laboratory of Magnetic Resonance and Atomic and Molecular\\
Physics, Wuhan Institute of Physics and Mathematics, Chinese Academy of\\
Science, Wuhan 430071, People's Republic of China }
\maketitle

\begin{abstract}
We propose a method to create two-mode squeezed states and their
superposition in the center-of-mass mode and breathing mode of two-trapped
ions. Each ion is illuminated simultaneously by two standing waves. One of
the fields is tuned to excite resonantly and simultaneously both upper
sidebands of the two normal modes, while the other field tuned to the
corresponding lower sidebands.
\end{abstract}

\pacs{PACS: 42.50.Vk, 32.80.Pj\\
Keywords: Two-mode squeezed state; Center-of-mass mode; Breathing mode}

\vskip 1cm

\narrowtext

The superposition principle in quantum mechanics enables quantum states to
have some interesting distinct characteristics than classical ones, such as
coherence, squeezing and quantum entanglement. In recent years, much efforts
have been devoted to the field of generating variety quantum states to test
the validity of quantum mechanical fundamental predictions[1,2]. In the
context of cavity QED, a number of schemes have been presented for the
generation of nonclassical light fields[3-5]. Experimentally, squeezed light
and sub-Poissonian light have been produced[6].

A trapped-ion system turns out to be an alternative candidate for realizing
quantum-state preparation. The quantized vibrational motion of ions in the
trap potential plays a role of boson mode. When the trapped ions interact
with classical laser fields, its internal and external degrees of freedom
are coupled via the exchange of momentum with the laser fields. This
property provides the possibility of generating various nonclassical states
in the vibrational motion of trapped ions by exciting ions with appropriate
laser fields. So far, proposals to prepare various nonclassical motional
states of a trapped ion, such as Fock states [7], squeezed state [8], even
and odd coherent states [9,10] have been made. More important, schemes for
the motional quantum-state engineering via Fock state superposition [11] and
coherent state superposition on a line or on a circle [12] have been
proposed, which allows one to approximate many quantum states [13] and
provides a new way for quantum-state generation. In addition, the study on
creating motional states of multiple trapped ions [14-16] has also been
started. Experimentally, motional Fock states, squeezed states, coherent
states [17] and Schr$\ddot{o}$dinger Cat states[18] for the center-of-mass
mode of a single trapped ion have been observed.

The two-mode squeezed states are very important in quantum optics, since
several devices produce light which is correlated at two frequencies.
Usually these frequencies are symmetrically placed at either side of a
carrier frequency. The squeezing exists not in the single modes but in the
correlated state formed by the two modes[19]. This kind of correlation
violate certain classical inequalities and can be employed to explain EPR
paradox[20]. More recently, this kind of correlation has also played a
leading role in the quantum teleportation of continuous variables [21]. In
addition, the superposition of two-mode squeezed states is also important,
because it can approximate a variety of two-mode entangled states with
different degrees of entanglement.

In reference [10], the author has assumed that an ion is trapped in a
two-dimensional isotropic harmonic potential and driven by four lasers, two
along $x$ direction and two along $y$ direction, tuned to excite resonantly
the upper and lower vibrational sidebands in the two directions
respectively. In that way, entangled two-mode coherent states in the
two-dimensional normal modes of one ion have been prepared. While in
reference [15], the authors have used four lasers to excite respectively the
both upper and lower sidebands in the center-of-mass and breathing modes of
N-trapped ions, the same vibrational states, but in two normal modes of
N-trapped ions, have been produced. In this letter, we provide a method to
create two-mode squeezed states and their superposition in the
center-of-mass and breathing modes of two trapped ions, which needs only two
lasers to excite simultaneously the both upper and lower sidebands of the
two normal modes.

Let us consider two two-level ions of mass $m$ trapped in a linear trap
which are strongly bounded in the $y$ and $z$ directions but weakly bounded
in a harmonic potential in the $x$ direction. The two ions are placed
symmetrically at either side of the origin of the $x$ axis and oscillate
around their equilibrium positions, $x_{10}=-d/2$, $x_{20}=d/2$. We denote
by $\widehat{X}=(\widehat{x}_{2}+\widehat{x}_{1})/2$, $\widehat{x}=(\widehat{%
x}_{2}-\widehat{x}_{1})/2$ the center-of-mass and breathing mode operators,
respectively. Both ions are simultaneously illuminated by two classical
homogeneous standing wave lasers $E_{I}^{(+)}=E_{0I}\cos (k_{I}x+\varphi
_{I})e^{-i\omega _{I}t}$ and $E_{II}^{(+)}=E_{0II}\cos (k_{II}x+\varphi
_{II})e^{-i\omega _{II}t}$, with same amplitudes $E_{0I}=E_{0II}=E_{0}$ and
same effective wavevectors $k_{I}=k_{II}=k$, but with different frequencies $%
\omega _{I}$ and $\omega _{II}$. Experimentally, there are two ways to
implement the two standing waves: One is to use two-photon stimulated-Raman
transitions for each of the fields[22]. And the other, for single photon
electric-dipole transitions, one can arrange appropriately two standing
waves, which have wavevectors $\overrightarrow{k}_{I}^{\prime }$ , $%
\overrightarrow{k}_{II}^{\prime }$ and frequencies $\omega _{I}$, $\omega
_{II}$, so that the projective waves onto $\widehat{x}$ direction have the
required effective wavevectors $k_{I}=\overrightarrow{k}_{I}^{\prime }\cdot 
\widehat{x}=k_{II}=\overrightarrow{k}_{II}^{\prime }\cdot \widehat{x}$. We
assume that each ion is located at the anti-nodes of both standing waves, so
that $\varphi _{I}=\varphi _{II}=0$. The Hamiltonian of this system can be
written as,

\begin{equation}
H=H_0+H_{int},
\end{equation}

\begin{equation}
H_{0}=\mu a^{+}a+\nu b^{+}b+\omega _{0}(\sigma _{z1}+\sigma _{z2})/2,
\end{equation}

\begin{equation}
H_{int}=\sum_{i=1}^{2}\frac{\Omega }{2}\sigma _{+i}[(e^{-i\omega
_{I}t}+e^{-i\omega _{II}t})\cos (k\widehat{x}_{i})+H.C.]
\end{equation}
where $\mu $ ($\nu $) and $a$ ($b$) are the frequency and annihilation
operator of the center-of-mass mode (breathing mode), $\omega _{0}$ is the
energy difference between the ground state $\left| g\right\rangle $ and the
long-lived metastable excited state $\left| e\right\rangle $ of each ion, $%
\sigma _{zi}$ and $\sigma _{+i}$ are Pauli operators describing the internal
states of $i$th ion, $\Omega $ is the Rabi frequency associated with both
standing waves. For simplicity, the Planck's constant is set $\hbar =1$.We
start by taking the frequencies of the two standing wave lasers to excite
resonantly the both upper sidebands and both lower sidebands of
center-of-mass and breathing modes, e.g.

\begin{equation}
\omega _{I}=\omega _{0}-(\mu +\nu ),\quad \omega \allowbreak _{I}=\omega
_{0}+(\mu +\nu ).
\end{equation}
If both the center-of-mass mode and breathing mode are cooled under
Lamb-Dicke limit, then we can expand the Hamiltonian up to the order terms
of $\eta ^{2}$ or $\eta _{r}^{2}$ which represent the lowest couplings
between internal and external degrees of freedom. Transforming the above
Hamiltonian to the interaction picture with respect to $H_{0}$ and making
use of rotating wave approximation, we have,

\begin{equation}
H_{int}^{I}=\Omega \eta \eta _{r}(ab+a^{+}b^{+})(\sigma _{x1}-\sigma _{x2})
\end{equation}
where $\sigma _{xi}$ is the $x$ component of Pauli operator describing the
internal state of $i$th ion, and $\eta =k\sqrt{1/4m\mu }$ and $\eta _{r}=k%
\sqrt{1/4m\nu }$ are the Lamb-Dicke parameters corresponding to the
center-of-mass and breathing modes.

It is worthwhile to point out that the Hamiltonian (5) can also be achieved
by employing two classical running waves under similar conditions. The
advantage of using standing waves rather than running waves is that all the
odd order terms in the expansion of Hamiltonian (3) are suppressed, so that,
contrary to running waves, the incorrectness from those terms, particularly
from the first and the third order terms, vanishes automatically. In this
sense, the use of standing waves may make the scheme more reliable than the
use of running waves. In addition, reference [23] advanced a method to
produce similar coupling as (5) between the two-dimensional modes of a
trapped ion and its internal states, but which needs more number of lasers.
Observing the commutation relations of the spin and boson operators, the
exact propagator for this Hamiltonian reduces to

\begin{eqnarray}
U &=&\frac{1}{4}\{[S^{+}(G)+S(G)]^{2}-\sigma _{x1}\sigma
_{x2}[S^{+}(G)-S(G)]^{2}  \nonumber \\
&&+(\sigma _{x1}-\sigma _{x2})[S^{+}(G)+S(G)][S^{+}(G)-S(G)]\}
\end{eqnarray}
where $G=-i\Omega \eta \eta _{r}t$ and $S(G)=\exp (G^{*}ab-Ga^{+}b^{+})$ is
the unitary two-mode squeezing operator.

Now we assume that, initially, both the center-of-mass mode and breathing
mode are cooled to ground states, and the internal state are prepared in a
superposition of ground and excited states (which can be realized by using
carrier transition), so that the whole initial state is

\begin{equation}
\left| \psi _{0}\right\rangle =\frac{1}{2}\left( \left| e\right\rangle
_{1}-\left| g\right\rangle _{1}\right) \left( \left| e\right\rangle
_{2}+\left| g\right\rangle _{2}\right) \left| 0\right\rangle _{c}\left|
0\right\rangle _{r},
\end{equation}
where $\left| 0\right\rangle _{c}$ and $\left| 0\right\rangle _{r}$ refer to
vacuum states of center-of-mass mode and breathing mode, respectively. By
illuminating both the two ions simultaneously with the two standing waves
discussed above, e.g., performing unitary transformation (6) on the initial
state, after a interaction time $t$, the system evolves as,

\begin{equation}
\left| \psi _{1}\right\rangle =\frac{1}{2}\left( \left| e\right\rangle
_{1}-\left| g\right\rangle _{1}\right) \left( \left| e\right\rangle
_{2}+\left| g\right\rangle _{2}\right) S(2G)\left| 00\right\rangle _{cr}.
\end{equation}
At this moment, the internal and external degrees of freedom are unentangled
and the motional state is then in a two-mode squeezed vacuum state,

\begin{equation}
\left| 0,0,G\right\rangle _{squ}=S(2G)\left| 00\right\rangle _{cr}.
\end{equation}
The degrees of squeezing is measured by the squeezing factor $2G$ which can
be enhanced by simply increasing the interaction time $t$.

The unitary transformation (6) possesses an interesting configuration,
which, for the initial state of (7), entangles the two kinds of external
motion, but does not entangle the internal states of the two ions,
furthermore, the internal and external degrees of freedom of the evolved
state are always unentangled at any time. These properties enable us to
produce the wanted two-mode squeezed state without any measurements on the
internal states of the two ions.

The unitary transformation (6) can also be used to produce superposition
states of two-mode squeezed states with different degrees of squeezing. For
this purpose, we firstly prepare the initial state of the whole system as,

\begin{equation}
\left| \varphi _{0}\right\rangle =N_{0}\left( \left| e\right\rangle
_{1}+p_{1}\left| g\right\rangle _{1}\right) \left( \left| e\right\rangle
_{2}-p_{2}\left| g\right\rangle _{2}\right) \left| 00\right\rangle _{cr},
\end{equation}
where $N_{0}=\left[ \left( 1+\left| p_{1}\right| ^{2}\right) \left( 1+\left|
p_{2}\right| ^{2}\right) \right] ^{-1/2}$ with complex parameters $p_{1}$
and $p_{2}$ being the controlling weights of the internal levels of the two
ions respectively. Then, we let the system experience a unitary evolution
governed by (6) for a duration $T$, followed by a measurement on the
internal states of the two ions. With no fluorescence being detected, the
conditioned state of the system reads

\begin{equation}
\left| \varphi _{1}\right\rangle \sim \prod_{i=1}^{2}\left[
(1-p_{i})S(G)+(1+p_{i})S(-G)\right] \left| ee\right\rangle _{12}\left|
00\right\rangle _{cr}.
\end{equation}
We now set the internal states of the two ions again being in a
superposition form similar to (10), but with weight factors $p_{3}$ and $%
p_{4}$ for the two ions respectively. After an interaction time $T$ governed
by the unitary evolution (6), we perform a measurement on each ion.
Repeating this procedure $m$ cycles with each cycle having identical
interaction time $T$ , if no fluorescence being detected in all cycles, with
probability $p=\prod_{i=1}^{2m}\frac{1}{4}\left( 1+\left| p_{i}\right|
^{2}\right) ^{-1}$, the final conditioned state for the vibratic motion of
the two-ion system is

\begin{eqnarray}
\left| \varphi _{m}\right\rangle &\sim &\prod_{i=1}^{2m}\left[
(1-p_{i})S(G)+(1+p_{i})S(-G)\right] \left| 00\right\rangle _{cr} \\
&=&\sum_{k=0}^{2m}C_{2m}^{k}S\left[ 2(k-m)G\right] \left| 00\right\rangle
_{cr},  \nonumber
\end{eqnarray}
with 
\begin{equation}
C_{2m}^{k}=\sum \prod_{i\in \{k\}}(1-p_{i})\prod_{j\in \{2m-k\}}(1+p_{j}),
\end{equation}
where $\left\{ k\right\} $ denotes a set that picking $k$ elements out of $%
2m $ natural numbers corresponding to $2m$ parameters $p_{i}$, and $\left\{
2m-k\right\} $ denotes the complementary set to $\left\{ k\right\} $, e.g.,
the set constituted by the residual elements. The first $\prod $ represents
the multiplication of $k$ factors of $\left( 1-p_{i}\right) $ with $i$ being
the element of set $\left\{ k\right\} $, and the second $\prod $ represents
the multiplication of $\left( 2m-k\right) $ factors of $\left(
1+p_{j}\right) $ with $j$ being the element of set $\left\{ 2m-k\right\} $.
The sum in (13) is for all possible different sets $\left\{ k\right\} $ that
formed by picking $k$ elements out of $2m$ natural numbers.

Eq.(12) is a superposition state which contains one vacuum state and $2m$
two-mode squeezed states with different squeezing factors. By adjusting the
superposition coefficients, it can approximate many types of entangled
two-mode states, which have different degrees of entanglement. This is also
one of the methods to produce two-mode entanglement states. Of course, in
order to produce required two-mode entanglement states, we are usually
demanded to solve all of $p_{i}$ from given superposition coefficients $%
C_{2m}^{k}$. The detail description of treating this problem can be found in
reference[16].

Besides producing above two-mode squeezed vacuum states and their
superposition, we can also produce more general two-mode squeezed states and
their superposition. In reference [16], we have let two trapped ions
experience simultaneously two running waves. By adjusting the frequencies of
the two running waves to be resonant with the first red and blue side-bands
of center-of-mass mode, in Lamb-Dicke Limits and under rotating wave
approximation conditions, we have obtained the unitary propagator,

\begin{equation}
U_{ce}=\frac{1}{4}\prod_{i=1}^{2}\left\{ [D^{+}(\beta _{c})+D(\beta
_{c})]-\sigma _{yi}[D^{+}(\beta _{c})-D(\beta _{c})]\right\} ,
\end{equation}
with $\beta _{c}=i\eta \Omega t$ and $D(\beta _{c})=\exp (\beta
_{c}a^{+}-\beta _{c}^{*}a)$ being displacement operator for center-of-mass
mode. Similarly, if we adjust the frequencies of the two running waves to be
resonant with the first red and blue side-bands of breathing mode rather
than center-of-mass mode, in the same way, we can obtain a propagator,

\begin{eqnarray}
U_{re} &=&\frac{1}{4}\{[D^{+}(\beta _{r})+D(\beta _{r})]^{2}-\sigma
_{y1}\sigma _{y2}[D^{+}(\beta _{r})-D(\beta _{r})]^{2} \\
&&+(\sigma _{y1}-\sigma _{y2})[D^{+}(\beta _{r})+D(\beta _{r})][D^{+}(\beta
_{r})-D(\beta _{r})]  \nonumber
\end{eqnarray}
with $\beta _{r}=i\eta _{r}\Omega t^{\prime }$ and $D(\beta _{r})=\exp
(\beta _{r}b^{+}-\beta _{r}^{*}b)$ being displacement operator for breathing
mode.

Now let us describe how to create a general two-mode squeezed state. After
the state (8) is created, we let the two ions experience the following
unitary carrier transition

\begin{equation}
U_{e}=\frac{1}{4}\left( 1-i\sigma _{x1}+i\sigma _{y1}+i\sigma _{z1}\right)
\left( 1-i\sigma _{x2}-i\sigma _{y2}-i\sigma _{z2}\right)
\end{equation}
to prepare the state of the whole system as,

\begin{equation}
\left| \psi _{2}\right\rangle =\frac{1}{2}\left[ \left| e\right\rangle
_{1}-i\left| g\right\rangle _{1}\right] \left[ \left| e\right\rangle
_{2}-i\left| g\right\rangle _{2}\right] S(2G)\left| 00\right\rangle _{cr}.
\end{equation}
By performing unitary transformation (14) on this state for a time $t$, it
evolves,

\begin{equation}
\left| \psi _{3}\right\rangle =\frac{1}{2}\left[ \left| e\right\rangle
_{1}-i\left| g\right\rangle _{1}\right] \left[ \left| e\right\rangle
_{2}-i\left| g\right\rangle _{2}\right] D(2\beta _{c})S(2G)\left|
00\right\rangle _{cr}.
\end{equation}
Then we only let the ion $1$ experience a carrier transition

\begin{equation}
U_{e}^{\prime }=-\sigma _{z1}
\end{equation}
to reprepare the state as, 
\begin{equation}
\left| \psi _{4}\right\rangle =\frac{1}{2}\left[ \left| e\right\rangle
_{1}+i\left| g\right\rangle _{1}\right] \left[ \left| e\right\rangle
_{2}-i\left| g\right\rangle _{2}\right] D(2\beta _{c})S(2G)\left|
00\right\rangle _{cr}.
\end{equation}
By performing unitary transformation (15) on it for a time $t^{\prime }$,
the final state of the whole system becomes

\begin{eqnarray}
\left| \psi _{5}\right\rangle &=&\frac{1}{2}\left[ \left| e\right\rangle
_{1}+i\left| g\right\rangle _{1}\right] \left[ \left| e\right\rangle
_{2}-i\left| g\right\rangle _{2}\right] \otimes \\
&&D(2\beta _{r})D(2\beta _{c})S(2G)\left| 00\right\rangle _{cr},  \nonumber
\end{eqnarray}
and the motional state is just a general two-mode squeezed state 
\begin{equation}
\left| \beta _{c},\beta _{r},G\right\rangle _{squ}=D(2\beta _{c})D(2\beta
_{r})S(2G)\left| 00\right\rangle _{cr}.
\end{equation}

Following this way produced the superposition of two-mode squeezed vacuum
states, we can also prepare the superposition of (22) type of states i.e. a
superposition of general two-mode squeezed states.

In conclusion, we have studied the properties of two trapped ions
interacting simultaneously with two standing waves, where each ion is
located at the anti-nodes of both lasers. In Lamb-Dicke limits and under
rotating wave approximation, by adjusting the frequencies of the two
standing waves to drive resonantly the both upper and lower sidebands of
center-of-mass and breathing modes, we have successfully realized the
preparation of two-mode squeezed states. We have also presented a procedure
to generate the superposition of several two-mode squeezed states with
different degrees of squeezing, by using a combination of the methods
developed here and previous papers.

This work was supported in part by the National Natural Science Foundation
(19734006,10075018), EYTF of the Educational Department of China, Chinese
Academy of Science, Hunan Province STF, and a special project of NSF of
China via Institute of Theoretical Physics, Academia Sinica.

\end{document}